\documentclass[%
 reprint,
superscriptaddress,
 amsmath,amssymb,
 aps, prl
]{revtex4-2}

\usepackage{graphicx}
\usepackage{dcolumn}
\usepackage{bm}

\newcommand{\eqlab}[1]{\label{eq:#1}}
\renewcommand{\eqref}[1]{Eq.~(\ref{eq:#1})}

\newcommand{\figref}[1]{Fig.~\ref{fig:#1}}

\newcommand{\figlab}[1]{\label{fig:#1}}
\newcommand{\tabref}[1]{Table~\ref{tab:#1}}

\newcommand{\tablab}[1]{\label{tab:#1}}
\newcommand{\equal}{\!=\!}
\newcommand{\minus}{\!-\!}
\newcommand{\plus}{\!+\!}
\newcommand{\ket}[1]{|{#1}\rangle}

\newcommand{\braket}[2]{\langle #1 | #2 \rangle}
\newcommand{\tauG}{\tau_{\!\scriptscriptstyle \mathcal{G}}}
\newcommand{\OmG}{\Omega_{\scriptscriptstyle \mathcal{G}}}
\newcommand{\Tin}{T_{\rm{in}}}
\newcommand{\Tstore}{T_{}}
\newcommand{\chill}{\chi^{\scriptscriptstyle (2)} }
\newcommand{\chilll}{\chi^{\scriptscriptstyle (3)} }

\begin{document}

\newcommand{\msub}[1]{\mbox{\scriptsize #1}}
\newcommand{\tauin}{\tau_{\msub{in}}}

\preprint{APS/123-QED}

\title{Controlled-phase Gate using Dynamically Coupled Cavities and Optical Nonlinearities}

\author{Mikkel Heuck}
 \email{mheuck@mit.edu}
\affiliation{DTU Fotonik, Technical University of Denmark, Building 343, 2800 Kgs. Lyngby, Denmark}%
\affiliation{ Department of Electrical Engineering and Computer Science, Massachusetts Institute of Technology,
77 Massachusetts Avenue, Cambridge, Massachusetts 02139, USA}%

\author{Kurt Jacobs}
\affiliation{U.S. Army Research Laboratory, Computational and Information Sciences Directorate, Adelphi, Maryland 20783, USA}%
\affiliation{Department of Physics, University of Massachusetts at Boston, Boston, MA 02125, USA}%
\affiliation{Hearne Institute for Theoretical Physics, Louisiana State University, Baton Rouge, LA 70803, USA}%

\author{Dirk R. Englund}%
\affiliation{ Department of Electrical Engineering and Computer Science, Massachusetts Institute of Technology,
77 Massachusetts Avenue, Cambridge, Massachusetts 02139, USA}%

\date{\today}

\begin{abstract}
We propose an architecture for a high-fidelity deterministic controlled-phase gate between two photonic qubits using bulk optical nonlinearities in near-term feasible photonic integrated circuits. The gate is enabled by converting travelling continuous-mode photons into stationary cavity modes using strong classical control fields that dynamically change the cavity-waveguide coupling rate. This process limits the fidelity degrading noise pointed out by Shapiro [J.\ Shapiro, Phys.\ Rev.\ A, 73, 2006] and Gea-Banacloche [J.\ Gea-Banacloche, Phys.\ Rev.\ A, 81, 2010]. We show that high-fidelity gates can be achieved with self-phase modulation in $\chi^{\scriptscriptstyle(3)}$ materials as well as second-harmonic generation in $\chi^{\scriptscriptstyle(2)}$ materials. The gate fidelity asymptotically approaches unity with increasing storage time for a fixed duration of the incident photon wave packet. 
Further, dynamically coupled cavities enable a trade-off between errors due to loss and wave packet distortions since loss does not affect the ability to emit wave packets with the same shape as the incoming photons. Our numerical results show that gates with $99\%$ fidelity are feasible with near-term improvements in cavity loss using LiNbO$_3$ or GaAs.
\end{abstract}

\maketitle


The quest for determinstic photon-photon logic gates has generally been hindered by the absence of sufficiently strong nonlinearities at optical frequencies. One possible solution is to use detection as an effective nonlinearity~\cite{Knill2001}, but two-qubit gates realized this way are probabilistic and require large resource overheads~\cite{Kok2007}. Even with large Kerr nonlinearities, Shapiro showed in 2006 that two-photon gates between traveling wave packets cannot achieve high fidelity~\cite{Shapiro2006}; this fundamental limit was further elucidated in more recent work~\cite{Gea-Banacloche2010, He2011, Xu2013, Dove2014}. Newer theoretical proposals have re-opened the discussion by showing that arbitrarily high fidelity is possible in certain limits~\cite{Ralph2015, Brod2016, Viswanathan2018}, but their implementation appears to be very complex.

Here, we introduce an approach relying on dynamically coupled cavities to provide a means for absorbing, storing, and re-emitting photons from a multimode nonlinear optical cavity. Classical control fields couple two cavity modes via nonlinear wave mixing with a time-dependent coupling rate determined by the amplitude and phase of the controls. As illustrated in~\figref{concept figure}a, incident wave packets couple to cavity mode $a$ (green mode) and are transferred to mode $b$ (blue mode) via their controlled coupling. Destructive interference between the directly reflected component of the incident wave packet and light leaking out from mode $a$ causes complete absorption by adjusting the controlled coupling to transfer population from mode $a$ to $b$ at the right rate. Photons are then stored in mode $b$, which is decoupled from the waveguide, and subsequent control fields release them through mode $a$. A similar cavity configuration was recently proposed to separate temporal modes of propagating pulses~\cite{Reddy2018}.
%
\begin{figure}[!h]  
\centering
   \includegraphics[width=8.0cm] {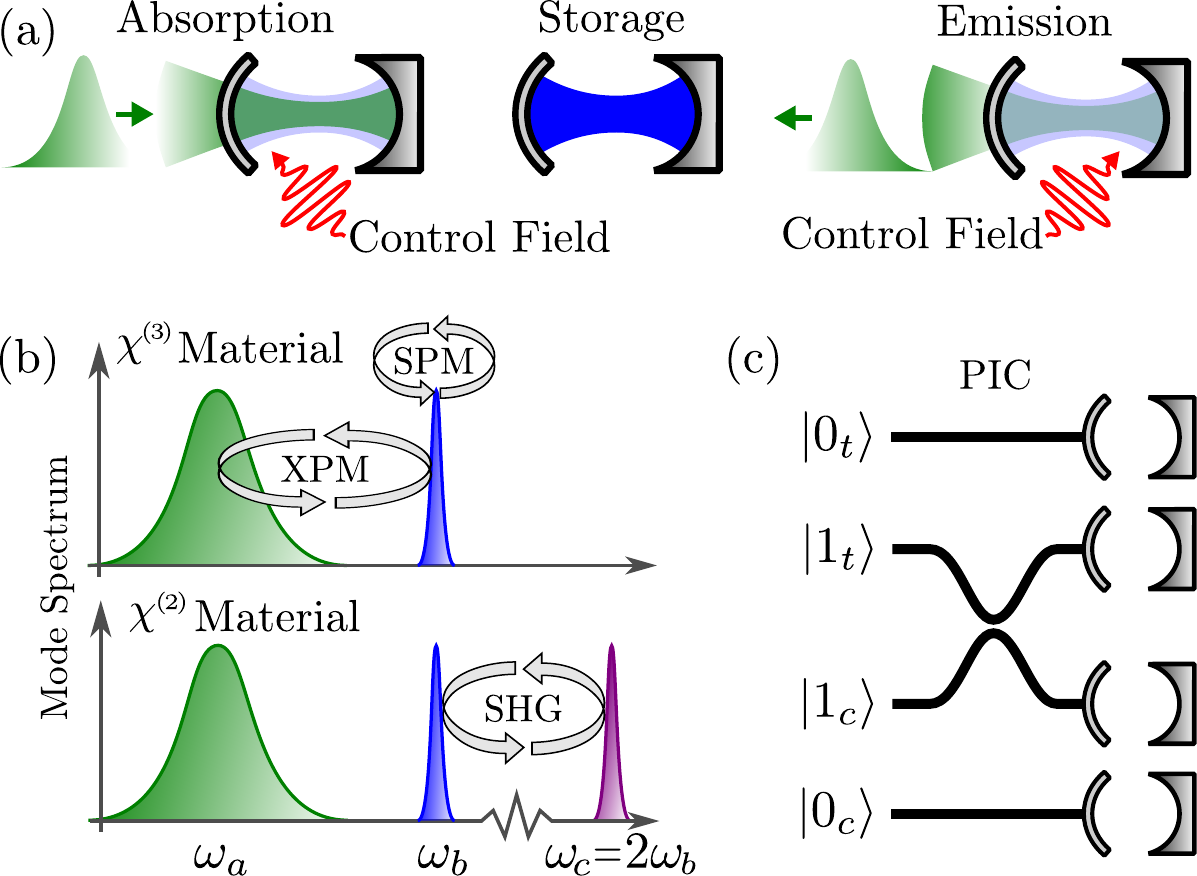} 
\caption{(a) Absorption, storage, and emission process. Mode $a$ (green) is waveguide-coupled and mode $b$ (blue) is decoupled. (b) Cavity mode-spectra showing the strongly waveguide-coupled mode with a correspondingly broad resonance (green) and decoupled modes (blue and purple) with narrow resonances. 
Photon-photon interactions for $\chilll$ and $\chill$ materials are also illustrated. XPM: Cross-phase modulation. SPM: Self-phase modulation. SHG: second-harmonic generation. (c) Photonic integrated circuit (PIC) that implements a controlled-phase gate on dual-rail encoded photonic qubits. }
\figlab{concept figure}
\end{figure} 

While stored in the decoupled cavity mode, photons are single-mode in the limit of zero intrinsic cavity loss. The fidelity limitations pointed out in Ref.~\cite{Shapiro2006} therefore do not apply to their interaction during this time. However, the control field that optimally absorbs and emits wave packets depends on the photon number when nonlinear interactions are present during the absorption and emission process. Since the same control field must be applied to any input state, it unavoidably introduces a finite amount of error consistent with Refs.~\cite{Shapiro2006, Gea-Banacloche2010, He2011, Xu2013, Dove2014}. Our numerical analysis reveals that this error scales favorably with the ratio between the storage time, $T$, and the duration of the input wave packet. The result is a scheme for high-fidelity photon-photon gates using bulk nonlinearities that should be feasible with near-term improvements in technology. \\

For an input state with two dual-rail encoded photonic qubits (see~\figref{concept figure}c), we denote the action of the gate by e.g. $\ket{0}_t\ket{1}_c \rightarrow \ket{\bar{0}}_t\ket{\bar{1}}_c$.  Single-photon input states like $\ket{1} \equal \!\int \!dt \xi_{\rm{in}}(t) \hat{w}^\dagger(t) \ket{\emptyset}$ are fully characterized by their wave packets, $\xi_{\rm{in}}(t)$, where normalization requires $\int \!|\xi_{\rm{in}}(t)|^2 dt\!\equal\! 1$ and $\hat{w}(t)$ is the continuous-time annihilation operator of the waveguide. Output wave packets are defined through $\ket{\bar{1}} \equal \!\int \!dt \xi_{\rm{out}}(t) \hat{w}^\dagger(t) \ket{\emptyset}$ or  $\ket{\overline{11}}\equal \int\!\int\! dt_mdt_n \xi_{\rm{out}}(t_m, t_n) \hat{w}^\dagger (t_m) \hat{w}^\dagger (t_n) \ket{\emptyset}$ corresponding to the input $\ket{1}_t\ket{1}_c\!\equiv\! \ket{11}$.
The controlled-phase operation corresponds to the phase requirement, $\arg(\braket{0}{\bar{0}})\equal \arg(\braket{1}{\bar{1}}) \equal (\arg(\braket{11}{\overline{11}})\plus \pi)/2$, and we define the one- and two-photon state fidelities as
\begin{subequations}\eqlab{fidelity}
\begin{align}
F_1 &\equal |\braket{1}{\bar{1}}|^2\equal |\!\!\int \!\xi_{\rm{out}}(t)^{\hspace{-0.2mm}*} \xi_{\rm{in}}(t\minus \Tstore\hspace{0.2mm}) dt|^2\\
    F_{\hspace{-0.2mm}11}\hspace{-0.2mm}&\equal \hspace{-0.2mm} \Big| \!\int \!\!\!\int \!\!\xi_{\rm{out}}\hspace{-0.3mm}(\hspace{-0.3mm}t_m \hspace{-0.3mm} ,t_n\hspace{-0.3mm})^{\!*} \xi_{\rm{in}}\hspace{-0.3mm}(\hspace{-0.3mm}t_n\!\!-\!\Tstore\hspace{0.2mm})\xi_{\rm{in}}\hspace{-0.3mm}(\hspace{-0.3mm}t_m\!\!-\!\Tstore\hspace{0.2mm})dt_ndt_m \hspace{-0.2mm} \Big|^{\hspace{-0.2mm}2} \!\!.
\end{align}
\end{subequations}
We consider loss by including a loss rate, $\gamma_L$, for all cavity modes. The output is therefore in a mixed state but we only calculate the dynamics of the zero-loss subspace where the output states above are not normalized and the fidelity in~\eqref{fidelity} is a lower bound~\cite{Heuck2019a}. 

To calculate the output wave packets, we use a Schr\"odinger-picture version of the established time-bin formalism~\cite{Scarani2002, Ciccarello2017, Gross2018}, which allows us to derive explicit equations of motion for the cavity states and input-output relations in terms of the cavity Fock basis. In the time-bin formulation the waveguide field is divided into $N$ time-bins of duration $\Delta t$, and the cavity interacts with the time bins one after the other. We refer to Ref.~\cite{Heuck2019a} for detailed derivations of all the equations of motion and input-output relations used here.  

We choose to calculate the controls so single-photon inputs are absorbed optimally into cavity mode $b$. We therefore only need to consider the linear part of the Hamiltonian without photon-photon interaction terms. For the $n^{\text{th}}$ time-step, in which the cavity interacts with time-bin $n$, this Hamiltonian in a rotating frame is
%
\begin{multline}\eqlab{Hn two mode}
    \hat{H}_n = i\hbar \sqrt{\frac{\gamma}{\Delta t}} \Big( \hat{a}^\dagger \hat{w}_n - \hat{a}\hat{w}_n^\dagger \Big) + \hbar \Big(\Lambda_n^{\!*} \hat{a}^\dagger\hat{b} + \Lambda_n \hat{b}^\dagger\hat{a}\Big), 
\end{multline} 
where $\hat{a}$ and $\hat{b}$ are annihilation operators for photons in cavity modes $a$ and $b$. The discrete-time annihilation operator of the waveguide in bin $n$ is $\hat{w}_n$. Its relation to the continuous-time operator is $\hat{w}(t_n)\equiv \hat{w}_n/\Delta t$ with $t_n\equal n\Delta t$ and $n\!\in\![0,N]$. $\gamma$ is the cavity-waveguide coupling rate of mode $a$ and $\Lambda_n$ is the coupling rate between modes $a$ and $b$, which is completely determined by the control fields. For $\chilll$ materials, $\Lambda_n$ arises from four wave mixing between two control fields and modes $a$ and $b$. To achieve energy-matching, the carrier frequencies obey the relation $\omega_2-\omega_1 \equal \omega_a-\omega_b$, where $\omega_1$ and $\omega_2$ are the carrier frequencies of the control fields. For $\chill$ materials, $\Lambda_n$ arises from three wave mixing between one control field and modes $a$ and $b$. In this case, their frequencies obey $\omega_p\equal \omega_a\minus\omega_b$, where $\omega_p$ is the carrier frequency of the control field.

In Ref.~\cite{Heuck2019a} we derived solutions for $\Lambda^{\!\scriptscriptstyle (k)}_n$ that enable complete absorption of a single photon with an arbitrary wave packet or emission of an arbitrary output, where $k$ refers to a $\chi^{\scriptscriptstyle (k)}$ material. The solutions differ due to cross-phase modulation imparted on cavity modes $a$ and $b$ by the control fields only in $\chilll$ materials.~\figref{results absorption}a shows an example of the absorption process with $\xi_{\rm{in}}$ being a Gaussian centered at $\Tin$ with temporal full-width-at-half-maximum $\tauG$ and spectral width $\OmG$. 
\begin{figure}[!h]  
\centering
   \includegraphics[height=3.3cm] {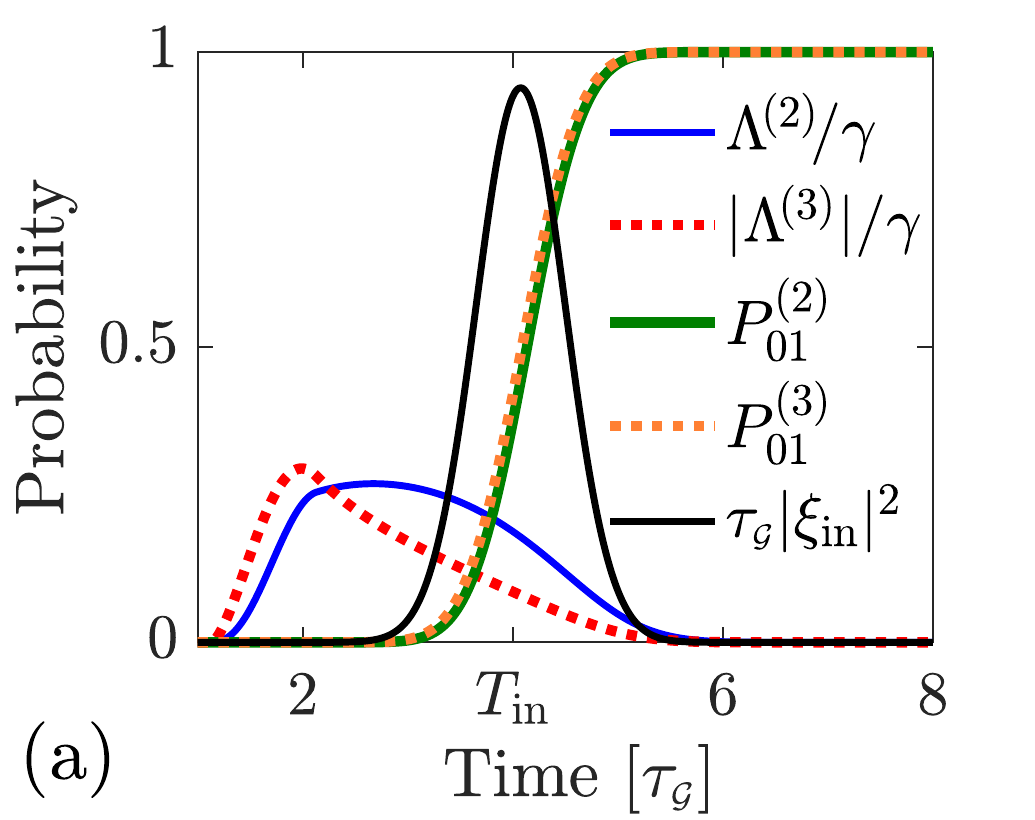}
   \includegraphics[height=3.3cm] {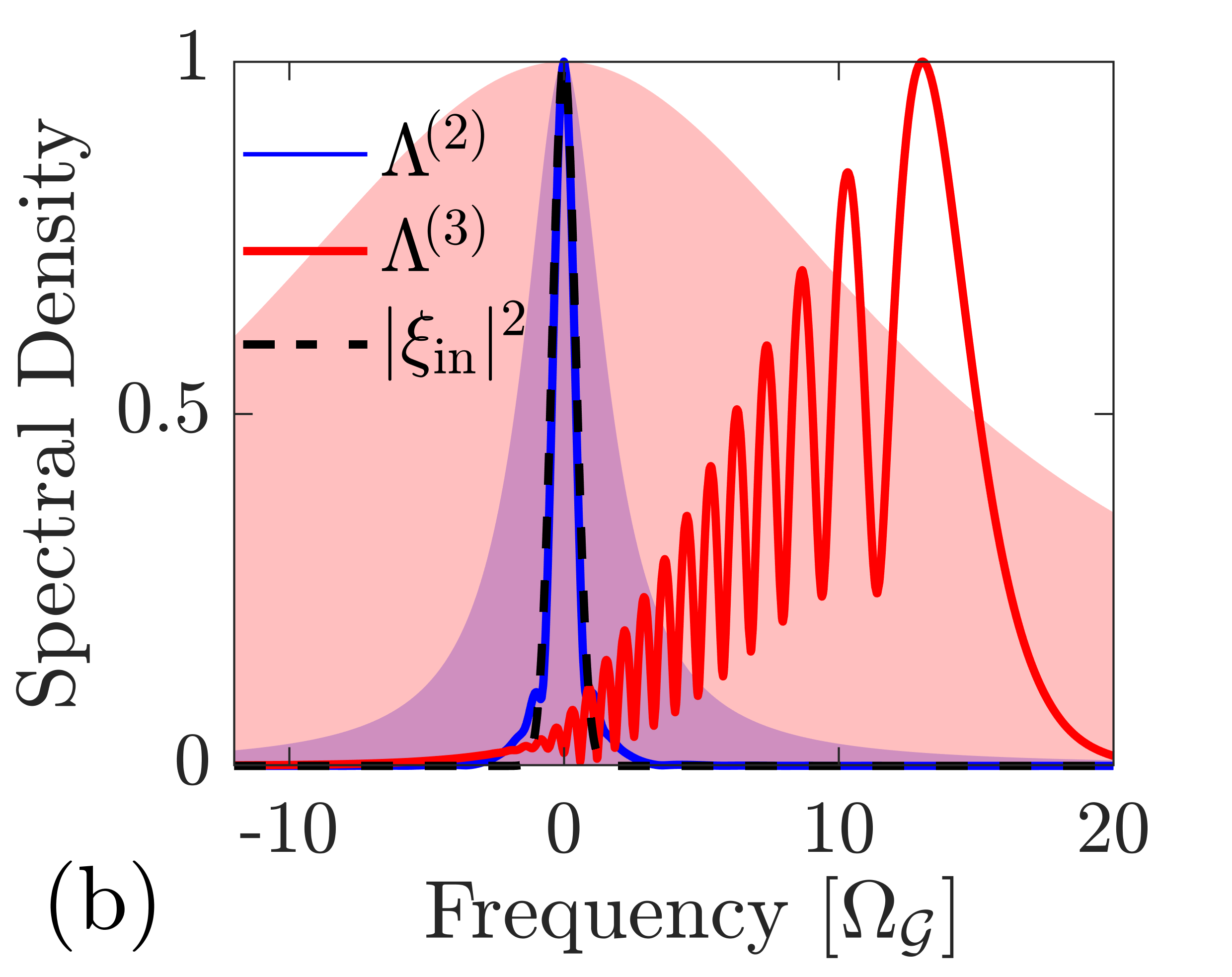}\vspace{3mm}
\caption{ (a) Absorption of a Gaussian wave packet including the solution of $\Lambda^{\!(k)}$ for both $\chill$ and $\chilll$ materials. (b) Fourier transformations of $\Lambda^{\!(k)}(t)$ from (a). The shaded areas plot Lorentzian resonances of mode $a$ with linewidths $\gamma/\OmG\equal 6$ for a $\chill$ material (blue) and $\gamma/\OmG\equal 30$  for a $\chilll$ material (red).  }
\figlab{results absorption}
\end{figure} 
The occupation probability of mode $b$ is $P_{01}^{\scriptscriptstyle (k)}$, where $k$ again refers to the order of the nonlinearity. $\Lambda^{\!\scriptscriptstyle(3)}$ has a time-dependent phase to compensate for the cross-phase modulation it induces on modes $a$ and $b$. This broadens and shifts its Fourier spectrum as seen in~\figref{results absorption}b. The absence of cross-phase modulation in $\chill$ materials also enables a similar absorption probability with a five times smaller coupling rate, $\gamma$, compared to $\chilll$ materials.

The probability of absorbing an incoming wave packet only depends on the ratio between mode $a$'s linewidth, $\gamma$, and the spectral width of the wave packet, $\OmG$.~\figref{results absorption b} plots the error in the one-photon state fidelity, $1\minus F_1$, for a Gaussian wave packet with a storage time of $T/\tauG\equal 14.4$. The different curves correspond to different loss rates, $\gamma_L$, which is assumed equal for all cavity modes.
\begin{figure}[!h]  
\centering
   \includegraphics[height=3.3cm] {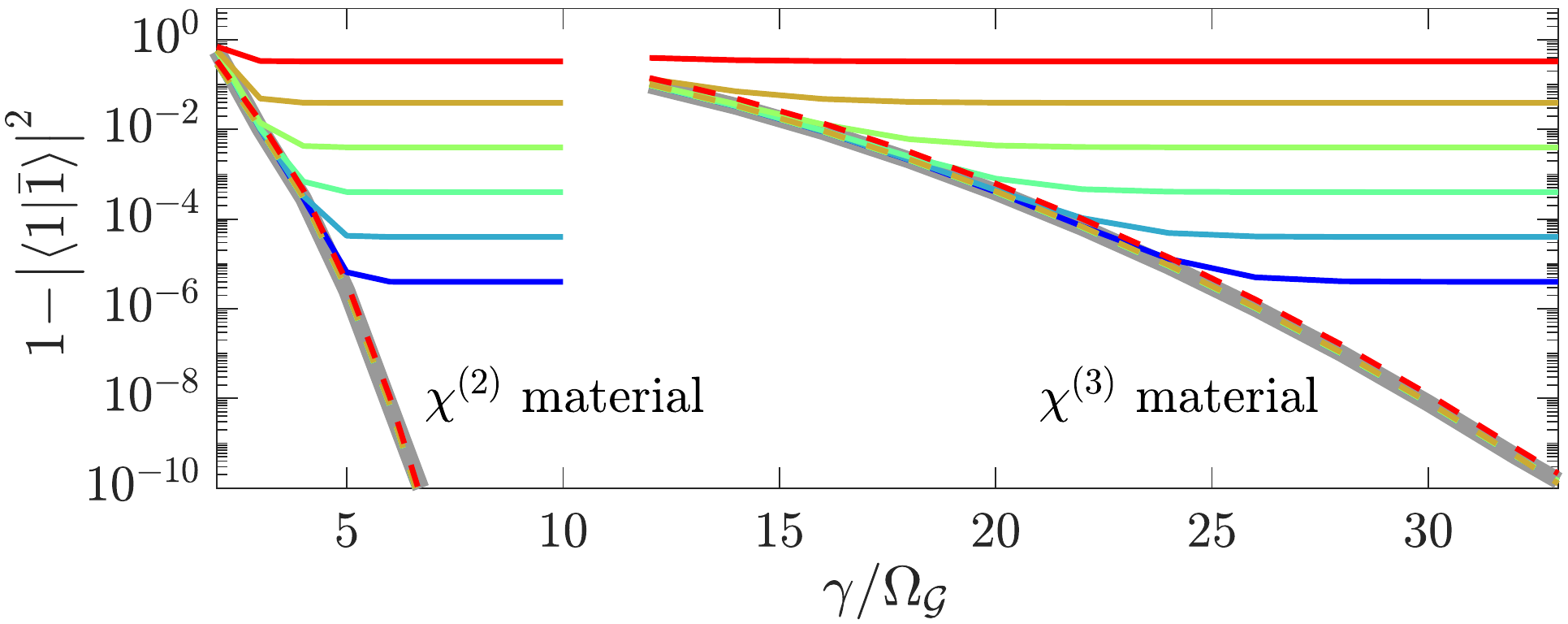}
\caption{Error in one-photon state fidelity, $1\minus F_1$, as a function of the linewidth of mode $a$ for different loss rates (solid lines). Gray corresponds to $\gamma_L/\OmG\equal 0$ while it increases from $10^{-7}$ (blue) to $10^{-2}$ (red) in steps of 10$\,$dB. Dashed lines plot the corresponding error in the conditional one-photon state fidelity, $1\minus \overline{F}_{\!1}$.   }
\figlab{results absorption b}
\end{figure} 
It shows how the error decreases much faster with increasing $\gamma/\OmG$ for $\chill$ materials than $\chilll$ materials due to the absence of cross-phase modulation. The curves flatten where the error becomes dominated by loss.~\figref{results absorption b} also plots the error in the conditional one-photon state fidelity defined by $\overline{F}_{\!1} \!\equiv\! F_1/\braket{\bar{1}}{\bar{1}}$. $\overline{F}_{\!1}$ may be understood as the probability of the input and output states being identical \emph{given} there was no loss because it corresponds to the fidelity calculated using the re-normalized state $\ket{\bar{1}}/\!\sqrt{\braket{\bar{1}}{\bar{1}}}$~\cite{Heuck2019a}.
The ideal scenario for lossy cavities is that the output wave packet is a scaled version of the input, $\xi_{\rm{out}}(t) \equal \sqrt{\eta} \xi_{\rm{in}}(t\minus T)$. For a given loss rate, $\gamma_L$, there is a corresponding value of $\eta$ from which $\Lambda^{\scriptscriptstyle (k)}_n$ is calculated to achieve $\xi_{\rm{out}}(t) \!\approx\! \sqrt{\eta} \xi_{\rm{in}}(t\minus T)$, see Ref.~\cite{Heuck2019a} for details. Since the conditional fidelity by definition is independent of the scaling factor $\eta$, we expect it to be negligibly dependent on loss so that $\overline{F}_{\!1}\!\approx\! F_1(\gamma_L\equal 0)$, which is confirmed in~\figref{results absorption b}. 
Thus, the photons will exhibit very high visibility quantum interference with other photons in Gaussian wave packets if they are not lost. For increasing loss, it is always possible to achieve such high visibility at the cost of a corresponding decrease in $\eta$. \\

The \emph{gate} fidelity is defined as the minimum \emph{state} fidelity for all input states~\cite{Chuang2011, Nysteen2017}. We can ensure that $F_1\!\approx\! 1$ if $\gamma/\OmG$ is large enough, which means that the gate fidelity is given by $F_{11}$. Below, we choose $\gamma/\OmG\equal 6$ for $\chill$ materials and $\gamma/\OmG\equal 30$ for $\chilll$ materials, which is seen to fulfill this requirement from~\figref{results absorption b}. The Hamiltonian in \eqref{Hn two mode} describes only the linear dynamics responsible for absorption and emission from the cavity. The nonlinear interactions responsible for the conditional phase shift are described by the Hamiltonians
\begin{subequations}\eqlab{H nonlinear chi2 chi3}
\begin{align}
    \hat{H}^{(2)}   &\!=\! \hbar\chi_2 \Big( \hat{c}\hat{b}^\dagger\hat{b}^\dagger + \hat{c}^\dagger\hat{b}\hat{b}\Big) \eqlab{H nonlinear chi2}\\
    \hat{H}^{(3)}   &\!=\! \hbar\chi_3 \Big( \hat{a}^\dagger\hat{a}\hat{b}^\dagger\hat{b} + \frac{\big( \hat{a}^\dagger\hat{a}\minus 1\big)\hat{a}^\dagger\hat{a} + \big(\hat{b}^\dagger\hat{b}\minus 1\big)\hat{b}^\dagger\hat{b}}{4} \Big), \eqlab{H nonlinear chi3}
\end{align}
\end{subequations}
where $\hat{c}$ is the annihilation operator for photons in mode $c$, see~\figref{concept figure}b. The nonlinear coupling rates are~\cite{Majumdar2013, Vernon2015b}
%
\begin{align}\eqlab{chi2 chi3 nonlinear coupling rate}
\chi_2 = \sqrt{\frac{\hbar\omega_b}{\epsilon_0}} \frac{\omega_b}{n^3} \frac{\chi^{(2)}}{\sqrt{ V_m^{(2)}}} ~~~ \text{and} ~~~ \chi_3 = \frac32 \frac{\hbar\overline{\omega}^2}{\overline{n}^4\epsilon_0}\frac{\chi^{(3)}}{V_m^{(3)}}, 
\end{align} 
where $\overline{\omega}^2\equal \!\sqrt{\omega_a\omega_b\omega_1\omega_2}$, $\overline{n}^2\equal \!\sqrt{n(\omega_a)n(\omega_b)n(\omega_1)n(\omega_2)}$, $\epsilon_0$ is the vacuum permittivity, $\chi^{\scriptscriptstyle (k)}$ is the $k^{\text{th}}$-order nonlinear susceptibility, and $V_m^{\scriptscriptstyle (k)}$ is the mode volume for $k^{\text{th}}$-order interactions. 
%
%
%
In $\chilll$ materials the conditional phase shift arises due to the self-phase modulation experienced by two photons while stored in mode $b$. In $\chill$ materials a $\pi$ phase shift occurs when two photons undergo one Rabi oscillation between mode $b$ and $c$, see~\figref{concept figure}b. The storage time is adjusted to ensure that the phase requirement is fulfilled in both cases.  \\

\figref{results chi-3}a shows the error, $1\minus F_{11}$, for a $\chilll$ material as a function of storage time for different values of the cavity loss rate, $\gamma_L$. 
\begin{figure}[!h]  
\centering
   \includegraphics[height=3.7cm] {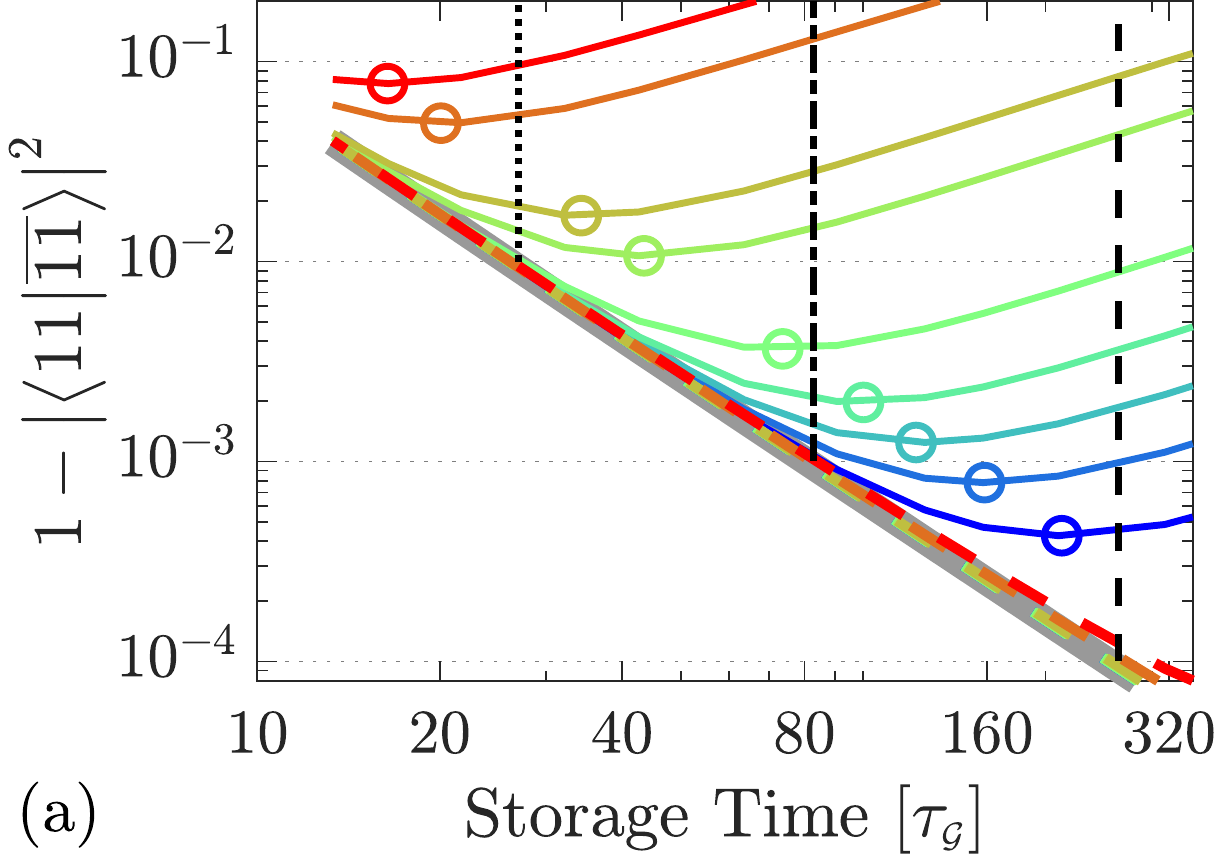}
   \includegraphics[height=3.7cm] {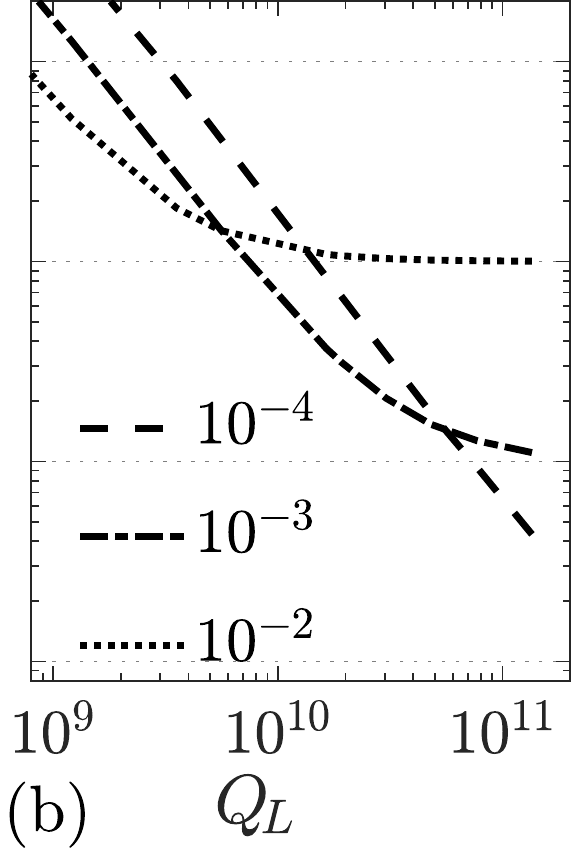}
\caption{(a) Plot of $1\minus F_{11}$ for a $\chilll$ material as a function of storage time, $\Tstore$, for different loss rates. Gray corresponds to $\gamma_L/\OmG\equal 0$ and it increases from $4\!\times\!10^{-7}$ (blue) to $10^{-3}$ (red). Dashed lines with the same color plot the corresponding values of the error in conditional fidelity, $1\minus \overline{F}_{\!11}$. (b) Plot of $1\minus F_{11}$ as a function of the intrinsic quality factor, $Q_L$, corresponding to the vertical cross-sections in (a). The legend shows the limiting values of the conditional fidelity, $1\minus \overline{F}_{11}$. Parameters: $\gamma/\OmG\equal 30$, $\chilll\equal 1.8\!\times\!10^{-19}\,$m$^2$/V$^2$~\cite{Moss2013}, $\lambda\equal 1550\,\text{nm}$, $\overline{n}\equal 3.4$, $V_m^{\scriptscriptstyle (3)}\equal 10^{-3}\,(\lambda/\overline{n})^3$.   }
\figlab{results chi-3}
\end{figure} 
Note that for each storage time, the nonlinear coupling rate, $\chi_3$, was chosen to achieve the phase requirement mentioned above. Without loss, the error scales as $1\minus F_{11} \!\propto\! 1/\Tstore^{\hspace{0.1mm}2.0}$ and 99$\,$\% fidelity is possible with $\Tstore/\tauG\!<\!30$. The dashed colored lines in~\figref{results chi-3}a plot the conditional fidelity, $\overline{F}_{\!11}\!\equiv\! F_{11}/\braket{\overline{11}}{\overline{11}}$. Note that $\overline{F}_{\!11}\!\approx\! F_{11}(\gamma_L\equal 0)$ as in~\figref{results absorption b}, which means that $1- \overline{F}_{\!11}$ may be understood as the error resulting from wave packet distortion alone, while $1- F_{11}$ additionally includes error from loss. 
Increasing the storage time (beyond the optimum indicated by circles in~\figref{results chi-3}a) allows for reduced wave packet distortions at the cost of increased loss, resulting in a trade-off between the two error mechanisms. 

\eqref{chi2 chi3 nonlinear coupling rate} may be used to convert the normalized loss rate, $\tilde{\gamma}_L\equal \gamma_L/\OmG$, into an intrinsic quality factor, $Q_L\equal\omega/\gamma_L$. We do this using the parameters listed in the caption of~\figref{results chi-3} for a silicon cavity with an ultra-small mode volume~\cite{Hu2016, Choi2017, Hu2018}.~\figref{results chi-3}b plots the error, $1\minus F_{11}$, as a function of $Q_L$ for the three vertical cross-sections in~\figref{results chi-3}a corresponding to three limiting values of the conditional fidelity. The error is dominated by loss where the curves are linear and becomes dominated by wave packet distortion where the curves saturate.\\


\figref{results chi-2}a shows the error, $1\minus F_{11}$, for a $\chill$ material as a function of storage time for different values of the cavity loss rate. Here, the nonlinear coupling rate, $\chi_2$, is adjusted for each $\Tstore$ to ensure that it corresponds to one Rabi oscillation of the SHG process.
\begin{figure}[!h] 
\centering
   \includegraphics[height=3.7cm] {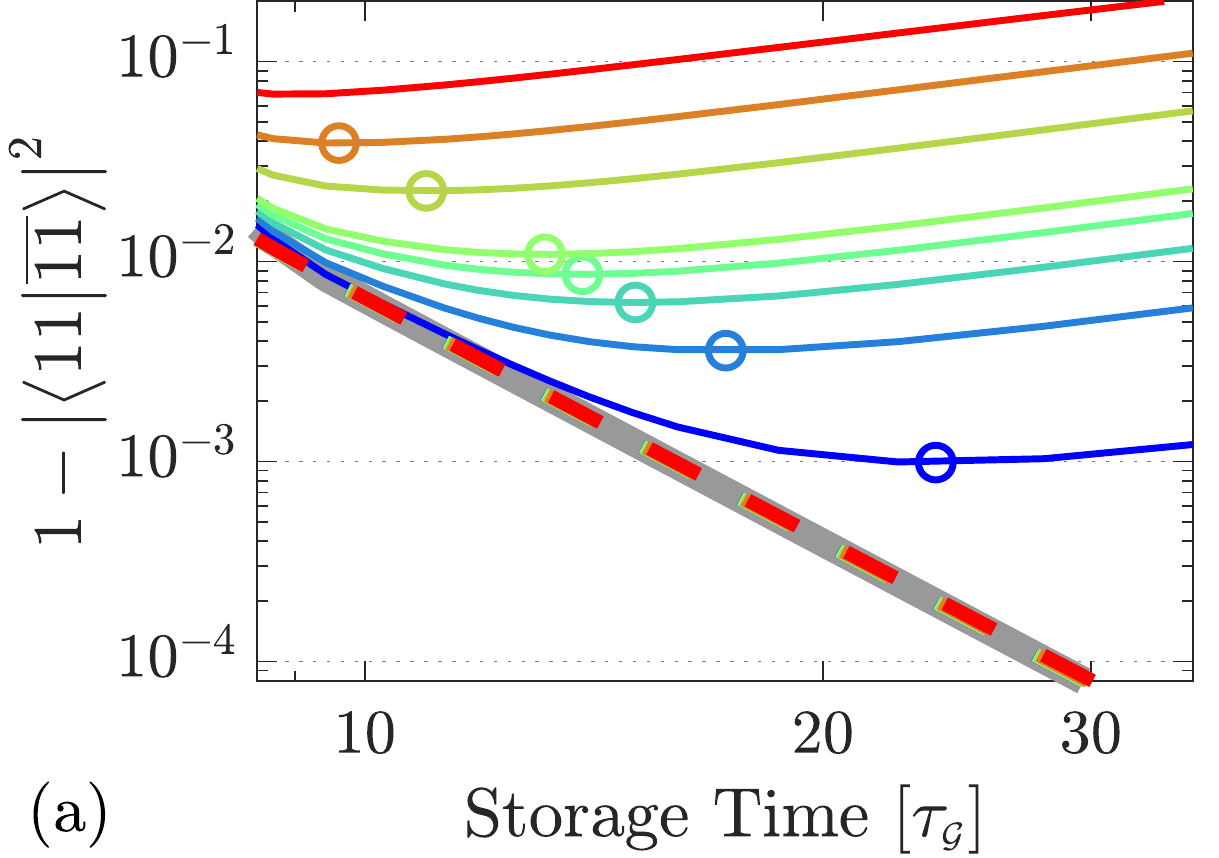}
   \includegraphics[height=3.7cm] {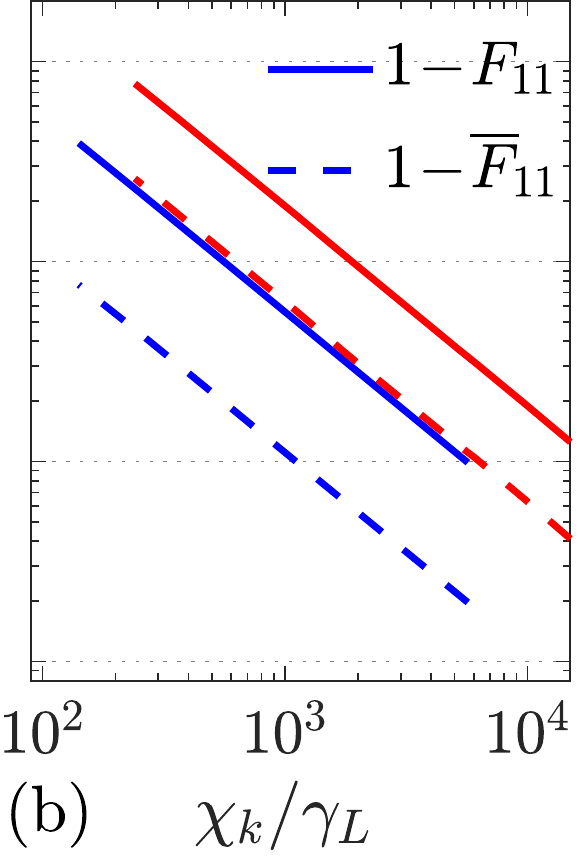}
\caption{(a) Plot of $1-F_{11}$ for second-order nonlinearity as a function of storage time. Gray corresponds to $\gamma_L/\OmG\equal 0$ and it increases from $10^{-5}$ (blue) to $2\!\times\!10^{-3}$ (red). Dashed lines with the same color plot the corresponding values of the error in conditional fidelity, $1\minus \overline{F}_{\!11}$. (b) Plot of the minimum error as a function of $\chi_k/\gamma_L$ corresponding to the circles in~\figref{results chi-3}a ($k\equal 3$, red) and~\figref{results chi-2}a ($k\equal 2$, blue). Dashed lines plot the corresponding values of the conditional fidelity. The slope of all curves are -1, demonstrating the relationship $1\minus F_{11} \equal C^{\scriptscriptstyle (k)} \gamma_L/\chi_k$, where $C^{\scriptscriptstyle (2)}\equal 5.5$ and $C^{\scriptscriptstyle (3)}\equal 18.7$. Note that (a) and (b) share the $y$-axis.}
\figlab{results chi-2}
\end{figure} 
The error-scaling is $1\minus F_{11}\!\propto\! 1/\Tstore^{\hspace{0.2mm}4.1}$, which is better than in~\figref{results chi-3}a since the photons only interact when they are both in mode $b$, while they interact during the entire absorption and emission process through both cross- and self-phase modulation for $\chilll$ materials. 
For the optimum choice of $\Tstore$ (indicated by circles in Figs.~\ref{fig:results chi-3}a and~\ref{fig:results chi-2}a), the error grows in proportion to the ratio between the loss rate and the nonlinear coupling, $1\minus F_{11} \equal C^{\scriptscriptstyle (k)} \gamma_L/\chi_k$, where $k$ again denotes the order of the nonlinear interaction.~\figref{results chi-2}b plots this relationship for both $\chill$ (blue) and $\chilll$ materials (red). It also shows that the conditional error, $1\minus \overline{F}_{\!11}$, follows the same relation but is 5.1 and 3.0 times smaller (dashed lines) for $\chill$ and $\chilll$ materials, respectively. The error may then be related to the quality factor and mode volume by 
\begin{align}\eqlab{error vs QL}
    1 \minus F_{11}^{(2)} \!= \mathcal{C}^{(2)} \frac{\sqrt{\widetilde{V}^{(2)}_m}}{ Q_L} ~~~\text{and} ~~~ 1 \minus F_{11}^{(3)} \!= \mathcal{C}^{(3)}  \frac{\widetilde{V}^{(3)}_m}{ Q_L},
\end{align}
where $V^{\scriptscriptstyle (k)}_m \equal \widetilde{V}^{\scriptscriptstyle (k)}_m (\lambda/n)^3$ and $\mathcal{C}^{\scriptscriptstyle (k)} \!\propto\! C^{\scriptscriptstyle (k)} / \chi^{\scriptscriptstyle (k)}$.~\tabref{results table} lists the values of $\mathcal{C}^{\scriptscriptstyle (k)}$ for the two most promising $\chill$ materials and the most common $\chilll$ material, silicon. The table also lists the required intrinsic quality factor to achieve a conditional fidelity of 99\% for an ultra-small mode volume, $\widetilde{V}_m^{(k)}\equal 10^{-3}$~\cite{Hu2016, Choi2017, Hu2018}, and a standard size for one-dimensional photonic crystal cavities, $\widetilde{V}_m^{(k)}\equal 0.5$.
\begin{table}[!h]
\renewcommand{\arraystretch}{1.7}
\vspace{3mm}
   \small
    \begin{tabular}{ |l|c|c|c|c|c|c|  }
    \hline
     {}&  \multicolumn{2}{c}{LiNbO$_3$} & \multicolumn{2}{|c|}{GaAs}& \multicolumn{2}{c|}{Si}\\
     \hline
     $\mathcal{C}^{(k)}$ &  \multicolumn{2}{c}{$5.0\!\times\!10^6$} & \multicolumn{2}{|c|}{$8.6\!\times\!10^6$} & \multicolumn{2}{c|}{$5.9\!\times\!10^{10}$}\\
     \hline
     $\widetilde{V}_m^{(k)}$  & $10^{-3}$  & 0.5   & $10^{-3}$  & 0.5 & $10^{-3}$  & 0.5 \\
     \hline
     $Q_L$  & $3\!\times\!10^6$  & $7\!\times\!10^7$    & $5\!\times\!10^6$  & $10^8$ & $2\!\times\!10^9$ & $10^{12}$\\
     \hline
    \end{tabular}
    \caption{Required values of the intrinsic quality factor to achieve a conditional fidelity of 99\% for three relevant materials. The corresponding values of $Q_L$ for a fidelity of 99\% are 5.1 times larger for $\chill$ materials and 3.0 times larger for $\chilll$ materials. Parameters: LiNbO$_3$: $\chi^{(2)}\equal 54\,$pm/V~\cite{Wang2017}, $\lambda\equal 1550\,\text{nm}$, $n\equal 2.1$. GaAs: $\chill\equal 270\,$pm/V~\cite{Skauli2002}, $\lambda\equal 3100\,$nm, $n\equal 3.5$. Si: $\chilll\equal 1.8\!\times\!10^{-19}\,$m$^2$/V$^2$~\cite{Moss2013}, $\lambda\equal 1550\,$nm, $n\equal 3.4$.  }
    \tablab{results table}
\end{table}
The numbers seem prohibitively large for $\chilll$ materials, but for $\chill$ they are close to state-of-the-art $Q$s in LiNbO$_3$, for which $10^7$ has been demonstrated in ring resonators both at telecom~\cite{Zhang_2017} and visible wavelengths~\cite{Desiatov2019} and $10^6$ in photonic crystal cavities~\cite{Liang2017a, Lin2019}. The large difference between $\mathcal{C}^{\scriptscriptstyle (2)}$ and $\mathcal{C}^{\scriptscriptstyle (3)}$ is primarily due to the difference in nonlinear coupling rates in~\eqref{chi2 chi3 nonlinear coupling rate}, but there is also a contribution from the difference between $C^{\scriptscriptstyle (2)}$ and $C^{\scriptscriptstyle (3)}$ in~\figref{results chi-2}b.\\ 

Our results show that dynamically coupled cavities offer a very promising approach to realize determinstic two-qubit gates between photons in a dual-rail encoding. With recent progress in nanofabrication of LiNbO$_3$ PICs~\cite{Zhang_2017, Liang2017a, Lin2019} and development of ultra-confined photonic crystal cavities~\cite{Hu2016, Choi2017, Hu2018} it appears that experimental demonstrations are within reach. Recent theoretical work also promises the required spectral properties of $\chill$ cavities~\cite{Minkov2019}. Noise sources not included in our analysis must also be investigated, including thermal noise-photons or photons generated by the control fields as well as higher-order nonlinear effects - in particular $\chilll$-effects in $\chill$ materials.

We acknowledge that experimental implementations remain very challenging, but hope it will stimulate the necessary near-term experimental advances to enable deterministic, high-fidelity photonic logic gates as well as extensions such as encoded logical qubits for quantum computing and one-way quantum repeaters. 

\textit{Acknowledgments:} The authors thank Joshua Combes and Jeffrey Shapiro for many useful discussions. This work was partly funded by the AFOSR program FA9550-16-1-0391, supervised by Gernot Pomrenke (D. E.), the MITRE Quantum Moonshot Program (M. H. and D. E.), the ARL DIRA ECI grant \emph{``Photonic Circuits for Compact (Room-temperature) Nodes for Quantum Networks"} (K. J.), and The Velux Foundations (M. H.).


%

\end{document}